\documentclass[10pt,conference]{IEEEtran}
\usepackage{cite}
\usepackage[tight,footnotesize]{subfigure}
\usepackage{graphicx}
\usepackage[cmex10]{amsmath}
\usepackage{amssymb}
\usepackage{mathtools}
\usepackage{multirow}
\usepackage{notoccite}
\usepackage{color} 
\usepackage{epsfig}
\usepackage{latexsym}
\usepackage{subfigure}
\usepackage{pstricks}
\usepackage{pdfpages}
\usepackage{epstopdf}
\DeclareGraphicsExtensions{.eps}
\usepackage{algorithmic}
\usepackage{algorithm}
\usepackage{multirow}
\usepackage{graphicx}
\usepackage{xurl}

\begin{document}
	\title{Integrated Access and Backhaul via Satellites}
	\author{\IEEEauthorblockN{Zaid Abdullah$^\dagger$, Steven Kisseleff$^\dagger$,  Eva Lagunas$^\dagger$, Vu Nguyen Ha$^\dagger$, Frank Zeppenfeldt$^\ddagger$, and Symeon Chatzinotas$^\dagger$ \\
			\\ $^\dagger$ Interdisciplinary Centre for Security, Reliability and Trust, University of Luxembourg, Luxembourg \\
   E-mails: \{zaid.abdullah, steven.kisseleff, eva.lagunas, vu-nguyen.ha, symeon.chatzinotas\}@uni.lu\\ 
			$^\ddagger$ European Space Agency (ESA), Noordwijk ZH, The Netherlands. E-mail: frank.zeppenfeldt@esa.int}}
	
\maketitle
\begin{abstract}
To allow flexible and cost-efficient network densification and deployment, the integrated access and backhaul (IAB) was recently standardized by the third generation partnership project (3GPP) as part of the fifth-generation new radio (5G-NR) networks. However, the current standardization only defines the IAB for the terrestrial domain, while non-terrestrial networks (NTNs) are yet to be considered for such standardization efforts. In this work, we motivate the use of IAB in NTNs, and we discuss the compatibility issues between the 3GPP specifications on IAB in 5G-NR and the satellite radio regulations. In addition, we identify the required adaptation from the 3GPP and/or satellite operators for realizing an NTN-enabled IAB operation. A case study is provided for a low earth orbit (LEO) satellite-enabled in-band IAB operation with orthogonal and non-orthogonal bandwidth allocation between access and backhauling, and under both time- and frequency-division duplex (TDD/FDD) transmission modes. Numerical results demonstrate the feasibility of IAB through satellites, and illustrate the superiority of FDD over TDD transmission. It is also shown that in the absence of precoding, non-orthogonal bandwidth allocation between the access and the backhaul can largely degrades the network throughput.  
\end{abstract}
\section{Introduction}
\subsection{\textit{Integrated Access and Backhaul: Motivation, Definition, and Network Architecture}}
Network densification refers to the deployment of additional radio access network (RAN) nodes in a given geographical area, to support wireless access to user equipments (UEs) with high quality of service (QoS) and low latencies. However, each additional RAN node would then need to backhaul its data to the core network (CN). Such backhauling can be performed either through the time-consuming and expensive installation of optical fiber cables, or over wireless channels (i.e. wireless backhauling). The latter approach is commonly known as \textit{integrated access and backhaul} (IAB), since the access for UEs and the backhaul between different RAN nodes are carried out over wireless channels. Such wireless backhauling plays a key role in allowing a cost-efficient and flexible network deployment~\cite{cudak2021integrated}, \cite{zhang2021survey}. 

The IAB concept was initially standardized by the third-generation partnership project (3GPP) over a decade ago in release 10~\cite{Rel10}. However, it is only recently that the IAB started to gain strong commercial interests thanks to the large available bandwidth in the fifth-generation new radio (5G-NR) spectrum (i.e. mmWave), and also the deployment of advanced antenna arrays with high link reliability known as massive multiple-input multiple-output. Two types of IAB have been defined by the 3GPP, in-band and out-of-band~\cite{report}. In out-of-band IAB, the access and backhaul links are allocated to different frequency bands, and hence, there is no interference between the access and the backhaul. In contrast, and when it comes to the in-band IAB, the two links (i.e. access and backhaul) should have at least partial frequency overlap, which could potentially lead to a better spectrum utilization. However, the 3GPP specifies that one should avoid the resultant interference when operating in an in-band IAB via separating the access and backhaul data streams in time, space, or frequency~\cite{report}.\footnote{Note that for the in-band IAB with frequency separation, the total bandwidth of the same frequency band is divided between the access and the backhaul links through frequency division multiplexing (FDM)~\cite{report}. Unlike the out-of-band IAB, where two separated frequency bands are utilized for the access and the backhauling.}   
 
\begin{figure}[t]
 \centering
       \includegraphics[scale=0.33]{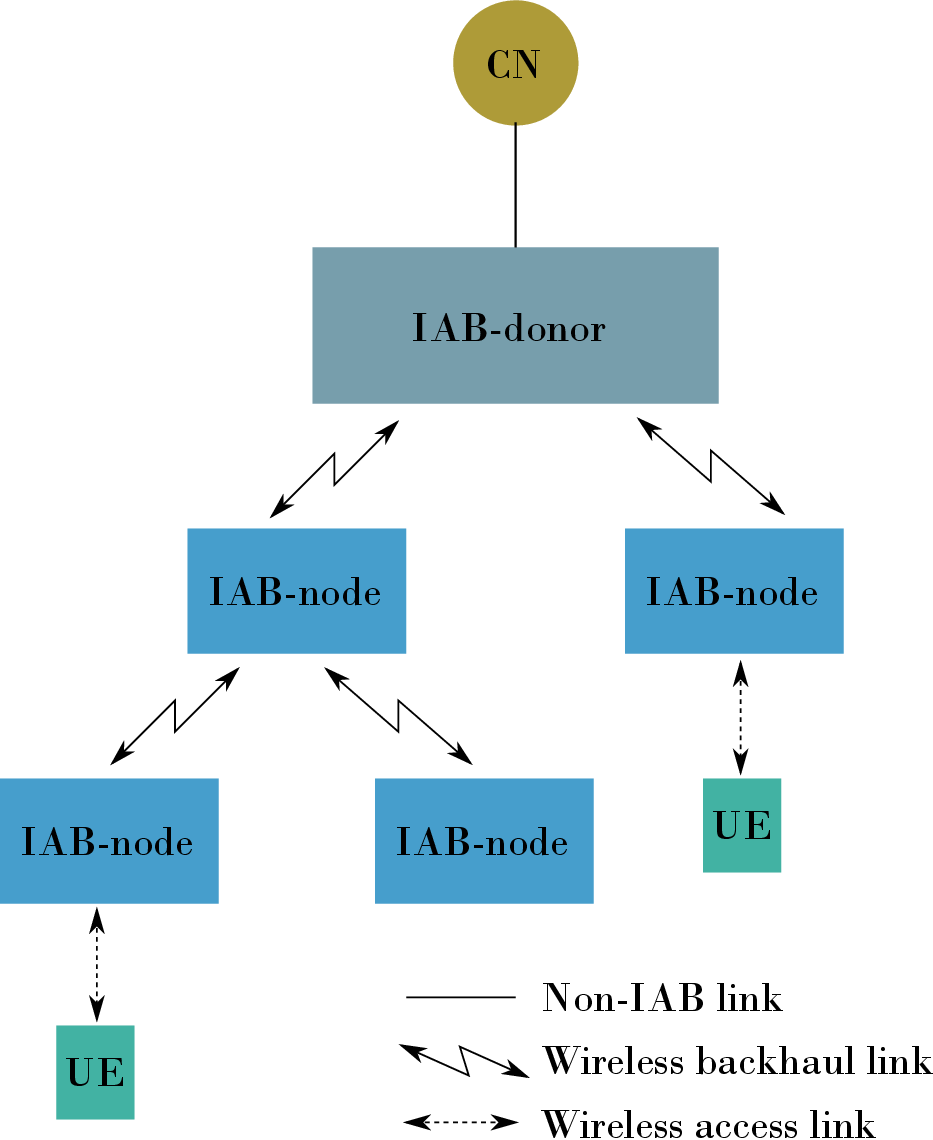} 
        \caption{NR-IAB network architecture.}
        \label{Fig1}
    \end{figure}

 According to the 3GPP specifications, an IAB network consists of an IAB donor and IAB nodes (see Fig.~\ref{Fig1})~\cite{report}. An IAB node is a RAN node that supports NR access links to UEs and NR backhaul links to other IAB nodes or to the IAB donor, while the IAB donor is a gNodeB (gNB) that provides network access to UEs through a network of access and backhaul links~\cite{Rel17_des}. Specifically, the IAB donor consists of one central unit (CU) with both user plane and control plane functionalities, and a minimum of one distributed unit (DU). The DU at the IAB donor is utilized to serve child nodes, which could either be IAB nodes or UEs, while the CU connects the IAB donor to the CN via a non-IAB link (such as optical-fiber cables). On the other hand, each IAB node comprises two functionalities, namely a mobile termination (MT) and a DU. An IAB node connects to child IAB nodes or UEs via its DU functionality, while the MT unit connects an IAB node to parent IAB nodes or to the IAB donor~\cite{report}. 
 
\subsection{\textit{IAB via Non-Terrestrial Networks}}
An aim of future wireless networks is to provide ubiquitous connectivity to everyone, everywhere, and at anytime. This includes serving users located in rural areas lacking modern telecommunications infrastructure, and those on ships and cruises in the middle of the oceans or on airplanes flying in the sky where it is impossible to install terrestrial RAN nodes. In such cases, providing IAB operation via non-terrestrial networks (NTNs) utilizing, for example, low/medium earth orbit (LEO/MEO) satellites seems a viable solution. Besides, NTN-enabled IAB can reduce the load of terrestrial networks when the terrestrial IAB nodes cannot cope with the ongoing demand. Furthermore, with the exponential increase in number of satellites orbiting the earth, IAB operation can become particularly important for efficient spectrum utilization in future NTNs.       

However, NTN-based IAB faces many crucial challenges compared to that in terrestrial networks. In particular, the 3GPP specifications and protocols for 5G-NR IAB in terrestrial networks are not always compatible with the satellite operation in terms of, for example, IAB-node mobility and transmission protocols as will be explained in detail in Section~\ref{compatibility}. Thus, such challenges need to be carefully addressed in order to assess the potential of IAB with NTNs in the era of 5G-NR and beyond. 
\subsection{Related Works}
In recent years, there has been a growing interest in utilizing NTNs for data backhauling. For example, the authors in~\cite{lagunas2017carrier} proposed a hybrid geostationary orbit (GEO) satellite-terrestrial backhaul network and solved the problem of carrier allocation for improved spectral efficiency of backhaul links. In~\cite{di2018data}, the authors proposed a terrestrial-satellite architecture to integrate ultra-dense LEO networks with terrestrial systems for data offloading. Aiming at maximizing the sum data rate, their results showed that integrated satellite-terrestrial networks can significantly outperform the non-integrated ones.  In addition, the work in~\cite{deng2020ultra} focused on the integrated LEO-satellite and terrestrial network for offloading. In particular, the work aimed at motivating data offloading between a satellite operator and a terrestrial operator for satellite-backhauled network access. The results showed that there exists an optimal LEO satellite density to balance the two operators’ utility and cost of data offloading. Finally, the work in~\cite{hu2020joint} considered the problem of user association and resource allocation for an integrated satellite-drone network. In their work, base stations (BSs) mounted on drones (DBSs) provided downlink connectivity to ground UEs whose demands cannot be satisfied by terrestrial BSs, while the satellite along with a set of terrestrial Macro BSs were utilized to provide resources for backhaul connectivity for both DBSs and small-cell BSs.
\subsection{Contribution}
The works~\cite{lagunas2017carrier, di2018data,  deng2020ultra} considered the use of NTNs only for backhauling, but with no radio access to UEs (i.e. no IAB operation). While the authors in \cite{hu2020joint} considered only a drone-based IAB with satellite backhauling (i.e. no satellite IAB). In our work, we focus on the satellite-based IAB operation which, to the best of our knowledge, has not been investigated before in the literature. In addition, one of our main contributions is that we highlight the challenges in realizing such systems from a practical point of view while taking into account the radio regulations and latest 3GPP specifications on IAB for 5G-NR networks. Moreover, a LEO-enabled IAB network architecture is presented, and resource optimization in terms of power and bandwidth is performed for such a network under both time- and frequency-division duplex (TDD/FDD) operations, assuming orthogonal and non-orthogonal bandwidth allocation between the access and the backhaul links. 

The rest of this paper is organized as follows. In Section~\ref{compatibility}, we address the main compatibility issues between the satellite operation and the 3GPP specifications on IAB. The system and channel models are then presented in Section~\ref{system model}. In Section~\ref{resource allocation}, we perform power and bandwidth allocation between access and backhauling to enhance the throughput. Section~\ref{results} presents the numerical results alongside their discussions. Finally, conclusions are drawn in Section~\ref{conclusions}.

\section{Compatibility Between Satellites IAB and 3GPP Specifications} \label{compatibility}
\subsection{Transmission Mode}
The 3GPP release 17 specifies that the NR IAB should operate in the TDD mode over the frequency bands in both frequency ranges (FR): FR1 (below $7.125$ GHz) and FR2 ($24.25 - 52.6$ GHz)~\cite{Rel17}. Indeed, for 5G-NR terrestrial networks, TDD is preferable since the RAN nodes (i.e. IAB donor and IAB nodes) should be equipped with large antenna arrays, particularly over FR2 frequency bands. Thus, TDD can greatly reduce the pilot signalling required to perform channel estimation compared to FDD transmission. However, most of today's satellites operate in the FDD mode in order to avoid the large differential delays of nodes in a wide satellie beam.\footnote{The only exception to that might be the IRIDIUM satellites which use the frequency range 1616 MHz to 1626.5 MHz (L-band) to communicate with ground users through TDD transmission. However, the FR1 frequency bands defined in the 3GPP release 17 for the NR IAB range from 2496 MHz to 5000 MHz \cite{Rel17}, and thus are incompatible with the IRIDIUM satellites despite the fact that both adopt the TDD mode.} As a result, there are two options in this regard to enable NTN-IAB, either (i) satellites need to operate in a TDD mode according to the 3GPP specifications on IAB, or (ii) the 3GPP needs to adopt an FDD-based IAB for the IAB donor and IAB nodes to enable NTN-IAB that is compatible with satellites operation. 

From a practical stand point, option (ii) seems more feasible especially that according to the 3GPP architecture description of next generation (NG) RAN, a gNB node (which is essentially an IAB donor) can support FDD, TDD, or dual mode operation \cite{Arch_desc}. In fact, the 3GPP has already made the first step towards this option by standardizing the NTN satellite operating bands n255 and n256 for UE access with FDD~\cite{UE_access}. 
\subsection{Mobile IAB Nodes}
 The 3GPP published a work item (WI) in September 2022 on mobile  IAB \cite{report1}, with the aim of a potential standardization in the ongoing 3GPP release 18. The WI specifies that mobile IAB (i.e. IAB nodes mounted on moving platforms such as vehicles) can be considered to provide 5G coverage/capacity enhancement for UEs. However, it highlights that in the 3GPP release 18, the functionality of mobile IAB nodes should not support child IAB nodes, and can only serve UEs.

 Such restrictions on IAB nodes' mobility can greatly limit the application of NTN-IAB, since all non-Geostationary satellites are essentially mobile IAB nodes due to their constant movement in the space. Therefore, in order to support flexible NTN-IAB, the 3GPP specifications in future releases would have to allow mobile NTN-IAB nodes to support wireless backhauling to terrestrial IAB nodes or to other NTN-IAB nodes via inter-satellite links in future terrestrial-space integrated networks.  
\section{System and Channel Models} \label{system model}
In this section, we first introduce the proposed network architecture for the satellite-enabled IAB. Subsequently, we provide a detailed description of the adopted channel model between the satellite and the ground segment. 
  \subsection{System Model}
\begin{figure}[t]
 \centering
       \includegraphics[width=8cm,height=4.5cm]{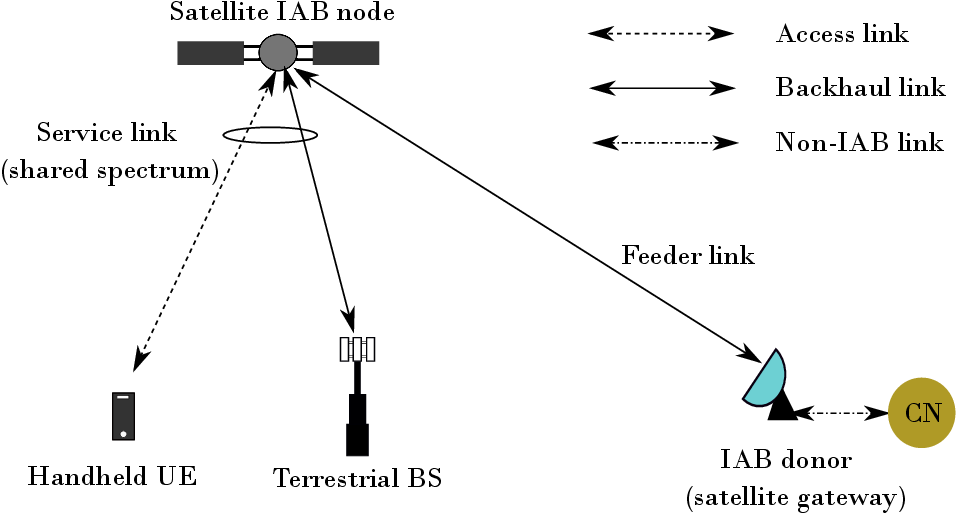} 
        \caption{The considered IAB network with a single satellite node.}
        \label{Fig2}
\end{figure}

We consider a network where a LEO satellite with IAB capabilities (hereafter referred to as sIAB) provides access to a ground UE with omnidirectional antenna (handheld) and backhaul to a terrestrial BS,\footnote{We assume that the BS is equipped with a very small aperture terminal (VSAT) antenna dish that is dedicated for communications with the satellites.}$^,$\footnote{Note that it is possible for the BS to reuse the spectrum and provide access to its own UEs in the area. However, our focus in this work is purely based on the IAB operation via the satellite node and not the terrestrial BS.} (see Fig.~\ref{Fig2}). The sIAB communicates with the IAB donor (satellite gateway) via the feeder link, which is assumed to be ideal and have a dedicated frequency band. Note that the communication between the gateway and the CN is typically performed via optical fibers, which is in line with the considered IAB architecture. The focus in this work will be on the IAB operation through the satellite forward service link, i.e. the link from the LEO sIAB node to both the UE and the BS, over the S-band frequency spectrum ($2-4$ GHz).\footnote{According to the 3GPP specifications on 5G-NR NTNs, a handheld UE can only communicate with the satellite via the S-band. In contrast, for a UE with a VSAT antenna dish, both the S-band and the Ka-band ($26-40$ GHz) can be used to facilitate the communication~\cite{UE_antenna}. }
  
We consider an in-band IAB with frequency division multiple access, where the total bandwidth of the same frequency band is split between the two links. Both orthogonal and overlapping bandwidth split will be considered as will be demonstrated in the next section. As such, the received signal at the UE and BS can, respectively, be given as:
\begin{subequations}
\begin{equation}
\small
y_{\text{UE}} = \sqrt{P_{\text{UE}}} h_{\text{UE}} s_{\text{UE}} + \varpi \sqrt{P_{\text{BS}}} h_{\text{UE}} s_{\text{BS}} + q_{\text{UE}} + n_{\text{UE}}, \label{y_UE}
\end{equation}
\begin{equation}
\small
y_{\text{BS}} = \sqrt{P_{\text{BS}}} h_{\text{BS}} s_{\text{BS}} + \varpi \sqrt{P_{\text{UE}}} h_{\text{BS}} s_{\text{UE}} + q_{\text{BS}} + n_{\text{BS}}, \label{y_BS}
\end{equation}
\end{subequations}where for $i\in\{\text{UE, BS}\}$, $P_i$ is the power allocated at the satellite for $i$th node in Watts, $h_i$ is the channel between the satellite and ground node~$i$, and $s_i$ is the information symbol intended for the same node satisfying $\mathbb E\{|s_i|^2\}=1$. The parameter $q_i$, which has a power spectral density (PSD) of $\kappa$, reflects the undesired interference due to imperfect synchronization and/or the satellite's out-of-band emissions that are hard to avoid especially for an in-band IAB scenario. Also, $n_i$ is the additive white Gaussian noise (AWGN) with zero mean and a PSD of $N_0$, while the parameter $\varpi$ in the second term of (\ref{y_UE}) and (\ref{y_BS}) accounts for the interference between access and backhaul links, such that we have $\varpi=0$ for orthogonal bandwidth allocation, and $\varpi=1$ for overlapping bandwidths.

Next, we introduce the channel model and channels gain.
\subsection{Channel Model}
We assume line-of-sight channels between the satellite and both the UE and the BS, which is a reasonable assumption given the high altitude of satellites. The free space path loss (PL) model is adopted to characterize and measure the gain of different communication channels. Specifically, for a ground node $i$ ($i\in\{\text{UE, BS}\}$), the corresponding channel gain is:
\begin{equation}
\small 
\beta_i = \frac{G_{\text{LEO}} G_i}{\text{PL}_i} ~ \psi(\theta_i),
\end{equation}where  $G_{\text{LEO}}$ and $G_i$ are the antenna gains at the satellite and a ground node $i$, respectively, $\text{PL}_i = (4\pi f_c d_i/c)^2$ is the free space PL between the satellite and node $i$ with $d_i$ being the corresponding communication distance, $f_c$ is the carrier frequency, and $c$ is the speed of light. Moreover, $\psi(\theta_i)$ accounts for the normalized antenna gain pattern at the satellite with respect to node $i$, such that
\begin{equation} \label{theta}
\small
\psi({\theta_i}) = 
\begin{cases}
1, \ \hspace{2.4cm} \text{if} \ \theta_i =0 , \\ \left|\frac{J_1(k.a.\sin{\theta_i})}{k.a.\sin{\theta_i}}\right|, \ \ \ \ \ \ \ \text{otherwise},
\end{cases}
\end{equation}where $\theta_i$ is the boresight angle from the sIAB node to ground node $i$, and $k = 2\pi f_c/c$. Also, the parameter $a$ in (\ref{theta}) represents the radius of the satellite's circular antenna aperture, while $J_1(\cdot)$ is the Bessel function of first kind and first order.

Next, we introduce the resource management schemes.

\section{Power and Bandwidth Allocation}\label{resource allocation}
In this section, we optimize the split of total power and bandwidth at the satellite between the access and backhaul links to enhance the throughput. In particular, a max-min optimization problem is considered which takes into account the deference in demands between the UE and the BS. Both TDD and FDD transmission modes are considered. 

We start by expressing the total throughput for the satellite forward service link (in bits/sec) as:
\begin{equation}
\small 
\mathcal R = \mathcal R_{A} + \mathcal R_{B},
\end{equation}where $\mathcal R_{A}$ and $\mathcal R_{B}$ are the access (at UE) and backhaul (at BS) achievable rates, respectively, given as:
\begin{subequations}
\small
\begin{equation}
\small 
\mathcal R_A = \alpha_o W_{{A}} \log_2\left(1 + \frac{P_{\text{UE}}\beta_{\text{UE}}}{(N_0 + \kappa)W_{{A}} + \varpi \alpha_1 \frac{P_{\text{BS}}\beta_{\text{UE}} w_o  }{W_B}   }\right), \label{RA}
\end{equation}
\begin{equation}
\small 
\mathcal R_B = \alpha_o W_{{B}}\log_2\left(1 + \frac{P_{\text{BS}}\beta_{\text{BS}}}{(N_0 + \kappa)W_{{B}}  + \varpi \alpha_1 \frac{P_{\text{UE}}\beta_{\text{BS}} w_o  }{W_A}    }\right), \label{RB}
\end{equation}
\end{subequations}where $W_{{A}}$ and $W_{{B}}$ are the bandwidths of access and backhaul links, respectively, and $w_o$ is the amount of bandwidth overlap between the two links. Particularly, the bandwidth for the access and the backhaul is subject to the constraint: 
{\small \begin{align}
& W_{{A}} + W_{{B}} \le \alpha_1 \left(W + w_o\right)
\end{align}}with $W$ being the total available bandwidth. Intuitively, we have $0 \le w_o \le W$, with $w_o=0$ means orthogonal bandwidth allocation, while $w_o = W$ reflects total overlap between the access and backhaul links.  In addition, $\alpha_o$ and $\alpha_1$ account for the utilized transmission mode such that:
\begin{equation}
\small 
\alpha_o = \begin{cases} 1, \hspace{1cm} \text{for FDD}, \\ \frac{1}{2}, \hspace{0.95cm} \text{for TDD,}
\end{cases} , \ \alpha_1 = \begin{cases} \frac{1}{2}, \hspace{0.95cm} \text{for FDD}, \\ 1, \hspace{1cm} \text{for TDD.}
\end{cases}
\end{equation}

Then, after introducing $\zeta$ as a slack variable, the following max-min optimization problem is formulated  
{\small \begin{align} \label{OP1}
	& \hspace{.1cm}\underset{\substack{W_{{B}}, W_{{A}},\\ P_{\text{BS}}, P_{\text{UE}}, \zeta}}{\text{maximize}} \hspace{1cm} \zeta \\ 
	&\hspace{.1cm}\text{subject to} \nonumber \\ 
	& \hspace{.1cm} P_{\text{UE}} + P_{\text{BS}} \le P, \label{1a} \\ & 
 \hspace{.1cm} W_{{A}} + W_{{B}} \le \alpha_1 \left(W + w_o\right), \label{1b} \\ &  \hspace{.1cm} W_i \le \alpha_1 W, \hspace{0.75cm} i \in\{A, B\}, \label{1c} \\ & \hspace{.1cm} W_i \ge \alpha_1 w_o, \hspace{0.75cm} i\in\{A,B\}, \label{1d}\\ & \hspace{.1cm}  \alpha_o W_{{A}} \log_2\left(1 + \frac{P_{\text{UE}}\beta_{\text{UE}}}{(N_0 + \kappa)W_{{A}} + \varpi \alpha_1 \frac{P_{\text{BS}}\beta_{\text{UE}} w_o  }{W_B}   }\right) \ge \epsilon \zeta, \label{1e} \\ & \hspace{.1cm} \alpha_o W_{{B}}\log_2\left(1 + \frac{P_{\text{BS}}\beta_{\text{BS}}}{(N_0 + \kappa)W_{{B}}  + \varpi \alpha_1 \frac{P_{\text{UE}}\beta_{\text{BS}} w_o  }{W_A}    }\right) \ge \zeta, \label{1f} 
\end{align}}where $P$ is the total available power. Constraint (\ref{1a}) ensures that the total power budget will not be violated. In addition, constraints (\ref{1c}) and (\ref{1d}) ensure that the bandwidth allocated for the access or the backhaul cannot be larger than the total bandwidth, nor can it be less than the amount of bandwidth overlap. Finally, the weighting parameter $\epsilon$ in (\ref{1e}), which will be assigned a value of $\epsilon < 1$ in our work as will be seen in Section~\ref{results}, accounts for the fact that the access and backhaul links have different QoS requirements.

In the following, we solve the problem in (\ref{OP1}) under orthogonal and overlapping bandwidths of access and backhaul. 
\subsection{Orthogonal Bandwidth Allocation ($\varpi = w_o = 0$)}
When $w_o=0$, constraints (\ref{1e}) and (\ref{1f}) become concave with respect to both the power and the bandwidth optimization variables. In addition, constraints (\ref{1a}) and (\ref{1b}) are convex due to their linear forms, while constraint~(\ref{1c}) becomes redundant. Therefore, problem (\ref{OP1}) can be solved optimally in such case (i.e. when $w_o=0$) using software tools such as the CVX.\footnote{Note that although the CVX solver does not accept functions of the form $f(x)= x\log(\frac{1}{x})$, it can still optimally solve problem (\ref{OP1}) when $w_o = 0$ after simple manipulations on constraints (\ref{1e}) and (\ref{1f}) through the use of the function \textit{rel\textunderscore entr} in MATLAB, which, for any two variables $\{x,y\}>0$, can be defined as \textit{rel\textunderscore entr}$(x,y)=x\log(\frac{x}{y})$. }
\subsection{Non-orthogonal Bandwidth Allocation ($\varpi = 1, \ w_o>0$) }
When $w_o>0$, constraints (\ref{1e}) and (\ref{1f}) become highly non-convex due to the interference terms in their denominators. To overcome this challenge, we propose a \textit{particle swarm optimization} (PSO) algorithm, which is an iterative evolutionary approach known for its efficiency in finding near optimal solutions for various non-linear problems with low computational complexity~\cite{abdullah2022successive}. In the following, we detail the steps of the proposed PSO scheme.
\subsubsection{Initialization} The first step of the PSO is to generate a population of $N$ random solutions (known as \textit{particles}) at iteration $t = 0$, which we denote by $\boldsymbol F^{[0]} = [\boldsymbol f_1, \boldsymbol f_2, \cdots, \boldsymbol f_N]^T$, where $(\cdot)^T$ is the transpose operator. Each of the $N$ particles in $\boldsymbol F$ is a vector with a dimension equivalent to the number of optimization variables. For the considered problem, we aim to optimize four variables: $P_{\text{UE}}$, $P_{\text{BS}}$, $W_A$, and $W_B$. Thus, the population matrix has four columns, i.e. $\boldsymbol F\in\mathbb R^{N\times 4}$. The first column in $\boldsymbol F$ is linked to the values of $P_{\text {UE}}$, while the second column is associated with the values of $P_{\text {BS}}$. In addition, the third and fourth columns of $\boldsymbol F$ accounts for the different values of $W_A$ and $W_B$, respectively. The first two columns in $\boldsymbol F$ are initialized with random values between $0$ and $P$, while the last two columns are initialized with random values between $0$ and $\alpha_1(W+w_o)$.
\subsubsection{Normalization} Once a new population is generated or updated, we normalize the powers and bandwidths of each particle according to constraints (\ref{1a})-(\ref{1d}). Specifically, at iteration $t$ ($t \in \{0, 1, \cdots, T-1\}$), and denoting $[\boldsymbol F]_{i,j}$ as the element in $i$th row and $j$th column of $\boldsymbol F$, the power normalization of the $n$th particle is carried out as follows
\begin{equation}
\small 
\left[\boldsymbol F^{[t]}\right]_{n,j} = \frac{\left|\left[\boldsymbol F^{[t]}\right]_{n,j}\right|}{\left|\left[\boldsymbol F^{[t]}\right]_{n,1}\right| + \left|\left[\boldsymbol F^{[t]}\right]_{n,2}\right|} \times P,
\end{equation} for $j\in\{1,2\}$ and $n\in\{1, \cdots, N\}$. Similarly, the total bandwidth is normalized according to constraint (\ref{1b}) as:
\begin{equation}
\small 
\left[\boldsymbol F^{[t]}\right]_{n,l} = \frac{\left|\left[\boldsymbol F^{[t]}\right]_{n,l}\right|}{\left|\left[\boldsymbol F^{[t]}\right]_{n,3}\right| + \left|\left[\boldsymbol F^{[t]}\right]_{n,4}\right|} \times \alpha_1(W+w_o),
\end{equation} for $l\in\{3,4\}$ and $n\in\{1, \cdots, N\}$. In addition, the following bandwidth adjustments are performed on the entries of the last two columns of $\boldsymbol F$ to meet constraints (\ref{1c}) and (\ref{1d}):
\begin{equation} 
\small 
\left[\boldsymbol F^{[t]}\right]_{n,l} :=\begin{cases} \alpha_1 W, \hspace{1cm} \text{if}\ \left[\boldsymbol F^{[t]}\right]_{n,l} > \alpha_1 W, \\ \alpha_1 w_o, \hspace{0.95cm} \text{if}\ \left[\boldsymbol F^{[t]}\right]_{n,l} < \alpha_1 w_o.
\end{cases}
\end{equation}
\subsubsection{Fitness evaluation} Once the power and bandwidth constraints are met, we evaluate the fitness of each particle. Considering the $n$th particle (i.e. $\boldsymbol f_n$), its fitness, denoted by $\lambda_n$, is a quality measure of its solution, and can be defined as:
\begin{equation}
\small 
\lambda_n = \min\left\{\mathcal R_A(\boldsymbol f_n), \epsilon \mathcal R_B(\boldsymbol f_n)\right\}, 
\end{equation}where $\min\{x,y\}$ returns the minimum value between $x$ and $y$, while $\mathcal R_A(\boldsymbol f_n)$ and $\mathcal R_B(\boldsymbol f_n)$ are the obtained access and backhaul rates in (\ref{RA}) and (\ref{RB}) under the power and bandwidth values in $\boldsymbol f_n$. Analytically, this can be expressed as:
{\small \begin{align}
& \mathcal R_i(\boldsymbol f_n) =  \Big\{\mathcal R_i \Big| P_{\text {UE}} = \left[\boldsymbol f_n\right]_1, P_{\text {BS}} = \left[\boldsymbol f_n\right]_2, \\ \nonumber & \hspace{1.75cm} W_A = \left[\boldsymbol f_n\right]_3, W_B = \left[\boldsymbol f_n\right]_4\Big\},
    \end{align}}
where $i\in\{A,B\}$ and $[\boldsymbol f_n]_j$ is the $j$th element of $\boldsymbol f_n$. 
\subsubsection{Local best and global best} The next step of the PSO algorithm is to find the \textit{local best} and \textit{global best} solutions, which will be utilized to derive the population toward good solutions that are closer to the optimal points. Specifically, at any given iteration, the global best solution is the particle that has the highest fitness in the population at that iteration, which we denote as $\boldsymbol f_{\text{max}}$. On the other hand, the local best solution for the $n$th particle is the particle with the highest fitness among only the neighbouring particles of $\boldsymbol f_n$,\footnote{We adopt a ring topology for the population, thus, the nighbouring particles for $\boldsymbol f_1$ are $\boldsymbol f_2$ and $\boldsymbol f_N$ , while the neighbours of $\boldsymbol f_N$ are $\boldsymbol f_1$ and $\boldsymbol f_{N-1}$.} which we denote by $\boldsymbol l_n \in\mathbb R^{1\times 4}$.    
\subsubsection{Velocity and population updates}
Once the local and global best solutions are obtained, we evaluate the velocity matrix $\boldsymbol X\in\mathbb R^{N\times 4}$, which we initialize with zeros at the beginning of the algorithm (i.e. at $t=0$). At any arbitrary iteration $t \in \{0, \cdots, T-1\}$, the velocity is updated as follows:
{\small \begin{align}
\left[\boldsymbol X^{[t+1]}\right]_{n,m} = & \left[\boldsymbol X^{[t]}\right]_{n,m} + u_1 r_1 \left(\left[\boldsymbol l_n\right]_m - \left[\boldsymbol F^{[t]}\right]_{n, m}\right) \nonumber \\ & + u_2 r_2 \left(\left[\boldsymbol f_{\text{max}}\right]_m - \left[\boldsymbol F^{[t]}\right]_{n,m}\right),
\end{align}}where $m\in\{1, \cdots, 4\}$, $n\in\{1, \cdots, N\}$, $u_1$ and $u_2$ are learning factors, while $r_1$ and $r_2$ are random numbers drawn from a uniform distribution with values between $0$ and $1$. \newline The matrix $\boldsymbol X$ controls the direction of each particle toward convergence, and it is utilized to update the population as follows:
\begin{equation}
\small 
\boldsymbol F^{[t+1]} = \boldsymbol F^{[t]} + \mu \boldsymbol X^{[t+1]}
\end{equation}with $\mu$ being the inertia weight. 

In subsequent iterations, steps $(2)$ to $(5)$ will be repeated until a maximum number of iterations is reached. The particle that achieves the highest fitness value across all iterations ($T$) will be selected as the final solution. 
\newline The per-iteration computational complexity of the proposed PSO method has an order of $\mathcal O(4N)$.

\begin{table}[]
\centering
\caption{Simulation parameters.} \label{table1}
\begin{tabular}{|c|c|}
\hline 
\textbf{Parameter}             & \textbf{Value} \\ \hline
Total bandwidth ($W$)                & $40$ MHz         \\ \hline
Total transmit power ($P$)              & $40$ dBm - $50$ dBm  \\ \hline
Noise PSD ($N_0$)    & $-174$ dBm/Hz  \\ \hline
Satellite antenna gain ($G_{\text{LEO}}$) & $36$ dBi         \\ \hline
BS antenna gain ($G_{\text{BS}}$)               & $32.8$ dBi       \\ \hline
UE antenna gain ($G_{\text{UE}}$)               & $0$ dBi          \\ \hline
Carrier frequency ($f_c$)         & $2$ GHz (FR1)          \\ \hline
Interference PSD ($\kappa$)    & $-174$ dBm/Hz  \\ \hline
Radius of LEO antenna aperture ($a$) & $1.5$ m         \\ \hline
LEO satellite altitude         & $\{600, \ 1200\}$ Km   \\ \hline
Boresight angles from LEO    & $\{\theta_{\text{UE}}, \theta_{\text{BS}}\}= \{0^o, 0.8^o\}$       \\ \hline  PSO population size ($N$)    & $50$         \\ \hline  Maximum number of PSO iterations ($T$)    & $200$         \\ \hline PSO inertia weight ($\mu$)    & $0.01$         \\ \hline
PSO learning factors    & $u_1 = u_2 = 2$         \\ \hline Access rate weighting parameter ($\epsilon$)    & $\{0.05, 0.1, \ 0.2\}$         \\ \hline
\end{tabular}
\end{table}

\section{Results and Discussions}\label{results}

Here we present our simulation results on IAB. The different parameters adopted in our simulations are given in Table~\ref{table1}. 

\begin{figure}[t]
 \centering
       \includegraphics[width=6cm,height=4.8cm]{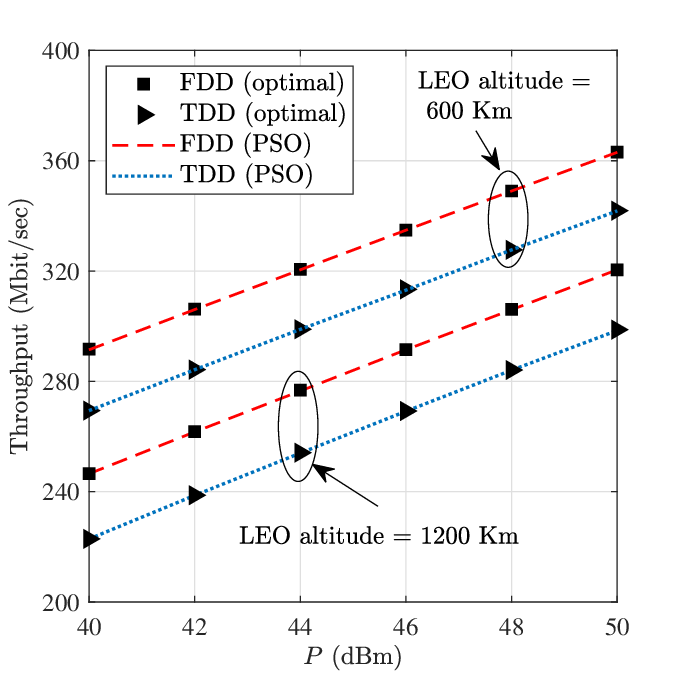} 
        \caption{Throughput vs transmit power when $\epsilon = 0.1$ and $w_o=0$.}
        \label{Fig3}
\end{figure}

Fig.~\ref{Fig3} illustrates the achievable throughput as a function of the transmit power levels at the sIAB node, under both TDD and FDD schemes and with different LEO altitudes. Orthogonal bandwidth allocation is performed with $w_o=0$. Clearly, higher transmit power levels lead to higher attainable throughputs. In addition, lower satellite altitude means a better link quality, and hence, higher achievable rates. Regardless of the satellite altitude and amount of transmit power, FDD brings better performance compared to the TDD scheme. This is the result of the higher noise levels accumalated over the wider signal bandwidth in TDD compared to FDD, which utilizes only part of the total bandwidth in each direction. In contrast, the TDD utilizes the full bandwidth in the forward and the reverse satellite service links, resulting in higher noise levels. The results also demonstrate the efficiency of the proposed PSO scheme, which shows an identical performance to the optimal solution obtained via the CVX solver.   

\begin{figure}[t]
 \centering
       \includegraphics[width=6cm,height=4.8cm]{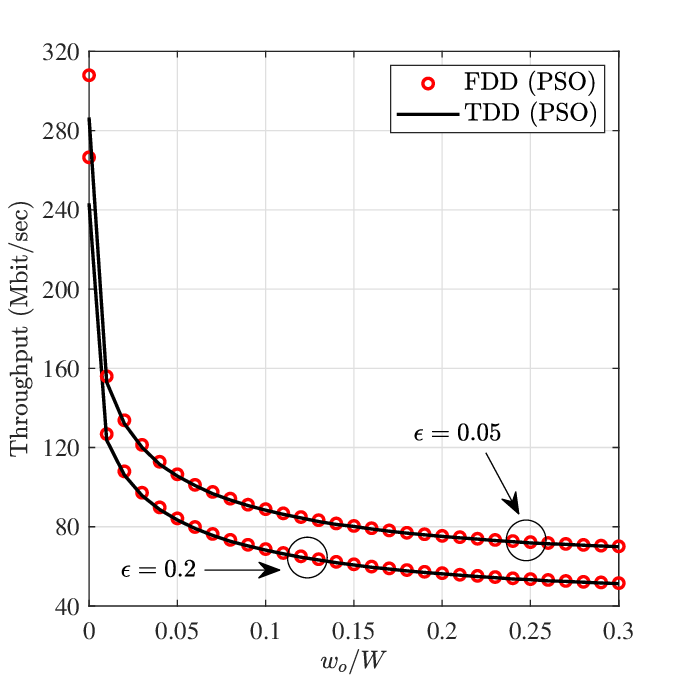} 
        \caption{Throughput vs amount of normalized bandwidth overlap $w_o/W$ when $P = 40$ dBm and LEO altitude $=600$ Km.}
        \label{Fig4}
\end{figure}

Fig.~\ref{Fig4} demonstrates the IAB rate performance vs. the amount of bandwidth overlap, and under different values of access weighting parameter ($\epsilon$). One can observe that the highest network throughput is attained when orthogonal bandwidths are assigned to the access and the backhaul (i.e. no overlap). This is in line with the 3GPP recommendations on separating the two in-band IAB data streams in time, frequency, or space~\cite{report}. The results also show the price to be paid when utilizing the satellite to provide access to a handheld UE. Specifically, one can notice that the total throughput (which takes into account both the access and backhaul) suffers from a large degradation as $\epsilon$ increases, where higher $\epsilon$ corresponds to a higher access rate. This shows that in order for the UE to enjoy a high QoS, the backhaul link will suffer dramatically. This is due to the link quality between the satellite and the handheld UE (which has an omnidirectional antenna) being much lower than that between the satellite and the BS, who has a high gain VSAT antenna. As a result, large amounts of power and/or bandwidth need to be allocated for the access link in order to compensate for the bad channel condition at the UE. Finally, it is also observed from Fig.~\ref{Fig4} that when $w_o>0$, FDD and TDD show almost identical performance. The reason is that in such cases, the interference due to the overlapped bandwidths is much larger than the noise power, and thus both schemes show a similar performance. 
\section{Concluding Remarks and Future Works} \label{conclusions}
Insights into the satellite IAB operation for future integrated terrestrial and NTNs were provided and discussed in this work. The main challenges and compatibility issues between satellites operation and terrestrial IAB standardization were pointed out based on the latest 3GPP technical reports and specifications. A case study was provided to demonstrate the IAB performance utilizing a LEO satellite, and with both TDD and FDD transmission modes. Numerical results showed that while it is feasible to provide IAB operation through satellites, the price to be paid can be high in terms of backhauling throughput. In addition, orthogonal bandwidth allocation demonstrated superior performance to that which allowed overlapping between the access and backhaul links. 

Our plans for future research directions include: (i) precoding at the satellite IAB node to deal with the interference in case of overlapping bandwidths, (ii) out-of-band IAB with satellites, and (iii) routing design through a constallation of satellite IAB nodes. 
\section*{Acknowledgment}
This work has been supported by the European Space Agency (ESA) funded activity Sat-IAB: Satellite and Integrated Access Backhaul - An Architectural Trade-Off (contract number 4000137968/22/UK/AL). The views of the authors of this paper do not necessarily reflect the views of ESA.
\bibliographystyle{IEEEtran}
\bibliography{SAT_IAB}	
\end{document}